\documentclass[usenatbib]{mn2e}
\usepackage[dvips]{graphicx}

\title[Planetary abundance from OGLE 2002 microlensing data.]{
The abundance of galactic planets from \mbox{OGLE-III} 2002 microlensing data.
}
\author[Snodgrass et al.]
{Colin~Snodgrass$^{1,2}$,
Keith~Horne$^{2}$,
Yiannis~Tsapras$^3$,\\
$^1$(C.Snodgrass@qub.ac.uk) APS Division, School of Physics, Queen's University Belfast, Northern Ireland BT7 1NN\\
$^2$(kdh1@st-and.ac.uk) School of Physics and Astronomy, University of St Andrews, Scotland KY16 9SS \\
$^3$(Y.Tsapras@qmul.ac.uk) School of Mathematical Sciences, Queen Mary, University of London, E1 4NS \\
}

\date{\today}

\newcommand{\Code}[1]{\begin{footnotesize}#1\end{footnotesize}}
\begin{document}

\maketitle

\begin{abstract}
From the 389 2002~OGLE-III observations of Galactic Bulge microlensing events we select 321 that are well described by a point-source point-lens lightcurve model. From this sample we identify $n=1$ event, 2002-BLG-055, which we regard as a strong planetary lensing candidate, and another, 2002-BLG-140, which is a possible candidate. If each of the 321 lens stars has 1 planet with a mass ratio $q=m/M=10^{-3}$ and orbit radius $a=R_{\rm E}$, the Einstein ring radius, analysis of detection efficiencies indicates that 14 planets should have been detectable with $\Delta\chi^2 > 25$. Assuming our candidate is due to planetary lensing, then the abundance of planets with $q=10^{-3}$ and $a=R_{\rm E}$ is $n_p \approx n/14$ = 7\%. Conversion to physical units (M$_{\rm Jup}$, and AU) gives the abundance of `cool Jupiters' ($m\approx{\rm M}_{\rm Jup}$,$a\approx4$~AU) per lens star as $n_p \approx n/5.5$ = 18\%. The detection probability scales roughly with $q$ and $(\Delta\chi^2)^{-1/2}$, and drops off from a peak at $a\approx4$~AU like a Gaussian with a dispersion of 0.4~dex.\end{abstract}

\begin{keywords}
Stars: planetary systems, extra-solar planets, microlensing --
Techniques: photometric --
\end{keywords}

\section{
Introduction
}
\subsection{
Microlensing
}\label{intro}
Gravitational lensing was first proposed by \citet{E} -- a consequence of his theory of general relativity, it is the effect of a massive body warping space-time and bending light around it, acting as a lens. This produces two or more images of background objects distorted around the `Einstein ring', which can be thought of as a scale radius ($R_{\rm E}$) of the lens and depends on its mass, $M$.  The angular radius of the ring is given by:
\begin{equation}
{\theta}_{\rm E} = \frac{R_{\rm E}}{D_L} = \sqrt{ \frac{D_{S}-D_L}{D_{S}D_{L}} \frac{4GM}{c^{2}}},
\end{equation}
where $D_L$ and $D_S$ are the observer -- lens and observer -- source distances respectively.  For objects of stellar mass, this radius is below the resolution limit of our telescopes, and we observe the phenomenon of microlensing.  In this case, we cannot resolve multiple images of the source, but detect an increase in brightness as the individual amplifications combine to produce a brighter stellar image.  As we observe a foreground (lens) star passing in front of a background (source) star, we see an increase in brightness up to a maximum amplitude $A_0$ at the time $t_0$ of closest approach, followed by a decrease.  This gives a symmetrical light curve around $t_0$, providing the timescale for the event is short enough for parallax effects to be negligible.  The timescale depends on the lens transverse velocity, relative to the observer-source line of sight, $v$, and is given by the length of time it takes the source to cross the Einstein ring of the lens:
\begin{equation}
t_{\rm E} = \frac{2R_{\rm E} }{v}.
\end{equation}

This depends on both mass and distances, through the dependence on $R_{\rm E}$.  $v$ is also indirectly influenced by distance, as the positions within the galaxy of the source and lens stars will give them average motions based on the orbital velocity at their galactic radius. For a survey looking toward the galactic bulge (i.e. $(D_S-D_L)/D_SD_L \sim$ kpc), a microlensing event due to a stellar mass ($M={\rm M}_{\odot}$) lens will last a few weeks. A Jupiter mass object will have $t_{\rm E} \approx$ 1 day, while an Earth would produce an effect which would be observable for only a few hours. 

If a stellar lens has a planetary companion, this will appear as a brief deviation from the stellar microlensing light curve, where the amplifications due to the star and planet combine \citep{MP,GL,BR}. These deviations should be observable, even for low mass objects, and therefore provide a method to detect extra-solar planets which is potentially more sensitive to small planets than radial velocity searches, given a programme of intensive monitoring with data taken many times per hour.

\subsection{
The OGLE project
}\label{OGLE}

The main difficulty with using gravitational microlensing to search for exoplanets is the low probability that any given source star will be lensed. This can be overcome by a monitoring programme which studies dense star fields and can detect a flux increase in any of the stars in the field.  The Optical Gravitational Lensing Experiment (OGLE) \citep{U} monitors $\sim$ 150 million stars in the galactic bulge, and alerts the astronomical community via Internet and e-mail alerts as soon as a lensing event is seen to begin. 

OGLE began in its current form, OGLE-III, in 2001. This version uses Difference Image Analysis (DIA) photometry \citep{W}, instead of the profile fitting (PSF) method used for the previous three seasons. This has substantially increased the detection rate of stellar microlensing events by OGLE -- there were 167 events discovered during the three years of \mbox{OGLE-II}, compared to 389 recorded during the 2002 season with \mbox{OGLE-III}.  

We use OGLE 2002 data to put firm upper limits on the population of planets in the bulge of our galaxy. Our method follows that of \citet{THK}, and is  described in Section \ref{lcf}, before Section \ref{2002data} looks at the results found for the 2002 data. The abundance of planets is discussed in Section \ref{ptot}.
 
\section{
Light curve fitting
}\label{lcf}
\subsection{PSPL Model}
\begin{figure}
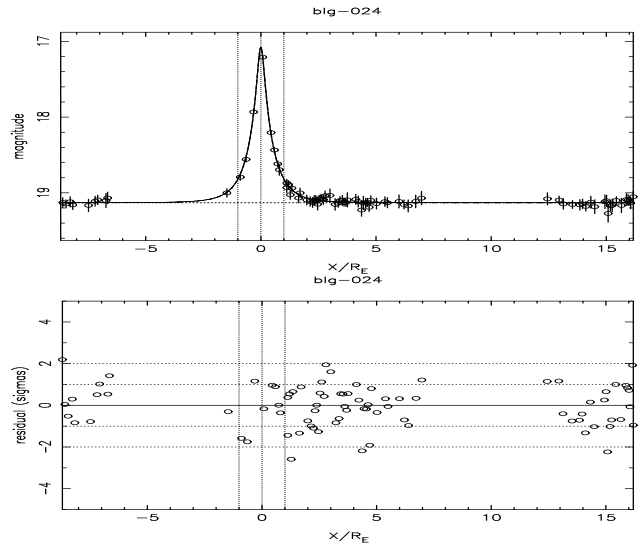

\begin{center}
\protect\label{mn024}
\includegraphics[angle=-90,width=0.47\textwidth,totalheight=0.2\textwidth]{fig1a.ps}\\
\includegraphics[angle=-90,width=0.47\textwidth, totalheight=0.2\textwidth]{fig1b.ps}
\caption{Light curve and normalised residuals for event 2002-BLG-024 -- an example of a good PSPL fit.}
\end{center}
\end{figure}

A Point-Source Point-Lens (PSPL) model of microlensing can describe the light curve entirely based on four parameters; $t_0$ and $t_{\rm E}$ as defined in Section \ref{intro}, the baseline (unlensed source) magnitude $I_0$, and the impact parameter $u_0$. The angular separation of the unlensed source and lens, measured in units of $\theta_E$, varies in time,
\begin{equation}
u(t) = \left[ u_{0}^2 + \left( \frac{2(t - t_0)}{t_{\rm E}}\right)^2 \right]^{1/2},
\end{equation}
and relates to the amplification of the signal, which governs the shape of the light curve with (Fig. \ref{mn024}) time, by:
\begin{equation}\label{At}
A(u(t))  = \frac{u^2 + 2}{u \sqrt{u^2 + 4}}.
\end{equation}
The impact parameter $u_0$ is the separation of the source and lens at closest approach, which is at $t_0$, and gives the maximum amplification $A_0=A(u_0)$.

\Code{PLENS} is a program which fits these parameters to microlensing data by minimising the  $\chi^2$ measure of `badness of fit'. In the case of microlensing, where $I_i$ and $\sigma_i$ are the reported magnitude and error bar for each data point,
\begin{equation}
\chi^2 = \sum_{i=1}^{N}\left(\frac{I_i - I(t_i)}{\sigma_i}\right)^2 .
\end{equation}
$I(t)$ represents the model light curve, and is given by:
\begin{equation}
I(t) = I_0 - 2.5 \log{\left( A(t) + b \right)}
\end{equation}
where the source magnification $A(t)$ is defined in Eqn.~\ref{At}, $I_0$ is the source star's baseline magnitude, and $b$ accounts for the additional flux from other stars that are blended with the source.

This method produces reasonable fits to the microlensing peak, however it was often found that the residuals in the flat (baseline) parts of the curve were larger than the photometric errors $\sigma_i$ reported for data points in this region. This was overcome by introducing a further fit parameter, $k_\sigma$, which scales the error bars. However this scaling often causes the error bars around the central peak to become much larger than the residuals, and would mask any interesting discrepancies. 

The scatter around the baseline can be explained by the fact that the fields observed are very densely populated, and so one further parameter was introduced: $\sigma_o$, the crowded field error. When $\sigma_0$ and $k_\sigma$ are used in conjunction they adjust the error bars to provide a much more accurate representation of the residuals. For $\sigma_0$ and $\sigma_i$ expressed in magnitudes, the adjusted error bars are then given by:
\begin{equation}
s_i = \sqrt{\frac{\sigma_0^2}{A^2(t)} + k_\sigma^2\sigma_i^2},
\end{equation}
and the best fit is found by maximising the likelihood criterion $L$, and is equivalent to minimising
\begin{equation}\label{likely}
-2\ln{L} = \chi^2 + 2\sum_{i=1}^{N} \ln{s_i}.
\end{equation}

There is one more fit parameter which \Code{PLENS} can adjust, the blending fraction $b$. This describes the amount of the baseline flux that is due to light from non-lensed stars in the crowded fields studied. 

\subsection{Planet detection}

\begin{figure}
\begin{center}
\protect\label{chi024}
\includegraphics[angle=-90,width=0.47\textwidth]{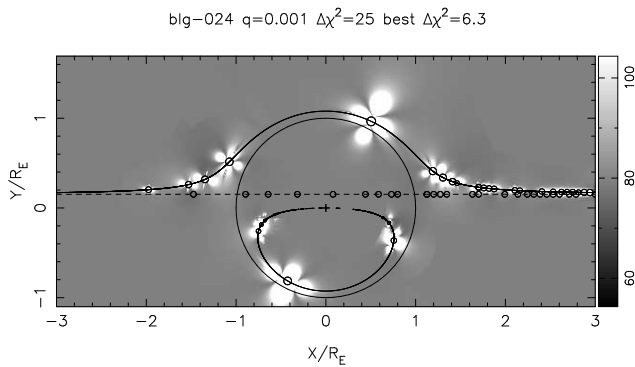}
\caption{$\Delta\chi^2$ map for event 2002-BLG-024.}
\end{center}
\end{figure}

$\Delta\chi^2$ maps (Fig. \ref{chi024}) were produced by adding a theoretical planet, with mass ratio $q=m/M$, at each point on a grid covering the lens plane. The resultant $\chi^2$ value is calculated, using a binary lens model, for each point. This results in a plot of `detection zones' around each data point, where the addition of a planet to the model gives $|\Delta\chi^2| > \Delta\chi^2_T$, where $\Delta\chi^2$ is the difference in $\chi^2$ between the binary and PSPL models, and $\Delta\chi^2_T$ is a threshold value of $\Delta\chi^2$. The plots show a white zone where the $\chi^2$ value is significantly increased, compared to a PSPL fit, by the addition of a planet. This means that the presence of a planet in these zones can be ruled out -- if there were a planet at that point, it would have significantly affected the data point at the centre of the zone. A black spot on the detection map would correspond to $\Delta\chi^2 < -\Delta\chi^2_T$, implying that the planet model improves the fit from the PSPL by a significant margin, which would be strong evidence for a planet. 

\begin{figure}
\begin{center}
\includegraphics[angle=270,width=0.47\textwidth]{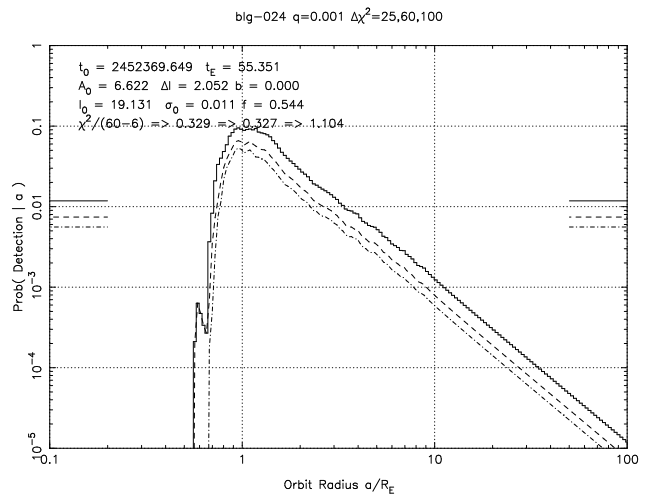}
\caption{Detection probability for event 2002-BLG-024. The curves show the probability of detection at $\Delta\chi^2$ thresholds of $\Delta\chi^2_T=25$,60 and 100.}
\protect\label{prb024}
\end{center}
\end{figure}

The probability of detecting a planet (Fig. \ref{prb024}) with a given mass ratio, $q$, and circular orbit of radius $a$, can be found by:
\begin{equation}
P({\rm det}|a,q) = \int P({\rm det}|x,y,q)P(x,y|a)dxdy.
\end{equation}
This equation simply sums, over all possible grid positions, the product of probability of detection at that position:
\begin{equation}
P({\rm det}|x,y,q)=\left\{ \begin{array}{l l}
  1& \textrm{if } |\Delta\chi^2| > \Delta\chi^2_T \\
  0& \textrm{otherwise,}  \\
\end{array}\right.
\end{equation}
and a geometrical term giving the probability that a randomly orientated orbit, of radius $a$, will put the planet at that point:
\begin{equation}
P(x,y|a)=\left\{ \begin{array}{l l}
\frac{1}{2\pi a \sqrt{a^2-r^2}} & \textrm{if }r=\sqrt{x^2+y^2}<a \\ 
\rule{0pt}{2.5ex}0 & \textrm{otherwise}.\\
\end{array}\right.
\end{equation}
The first term comes from the $\Delta\chi^2$ map -- it decides whether or not a planet at that point falls into a detection zone. As the detection zones are, in general, clustered around the Einstein ring (see Fig. \ref{chi024}), the highest detection probability is found at around $a \approx R_{\rm E}$. This is not surprising, as it is when a planetary lens interacts with a stellar lens at $R_{\rm E}$ that its effect is greatest.

\subsection{
Choice of detection threshold
}\label{residuals}
The total number of data points in the 321 OGLE 2002 events studied is 30582. A statistical treatment of this sample gives us the largest error expected from this much data (as a multiple of $\sigma$) using:
\begin{equation}
S_F (N) \approx 2.14\sqrt{\log{N}},
\end{equation}
assuming a Gaussian distribution of errors. For the 2002 data, with $N=30582$, the expected value of the largest residual is $4.53\sigma$, corresponding to a $\Delta\chi^2 \approx S^2_F \approx 20$. This implies that the choice of $\Delta\chi^2_T = 25$, corresponding to deviations from the PSPL fit of more than $5\sigma$, is appropriate for this data set -- any signal above this threshold should be larger than noise, and due to a real anomaly.

\subsection{
Analysis of photometric scatter
}\label{scatter}

\begin{figure}
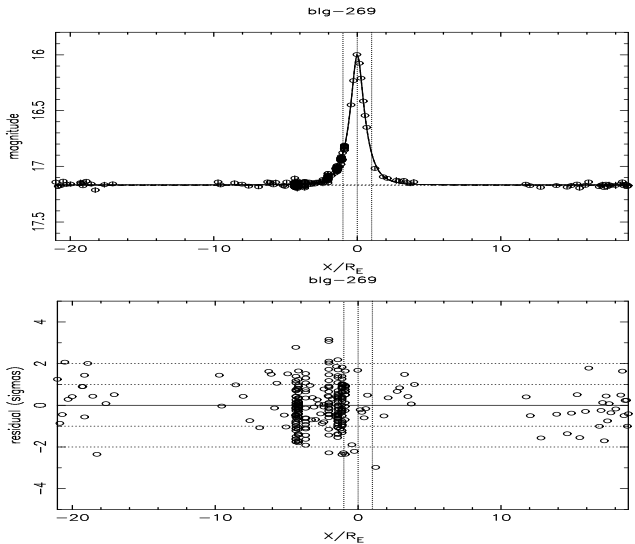

\begin{center}
\includegraphics[angle=-90,width=0.47\textwidth,totalheight=0.2\textwidth]{fig4a.ps}\\
\includegraphics[angle=-90,width=0.47\textwidth, totalheight=0.2\textwidth]{fig4b.ps}
\end{center}
\caption{2002-BLG-269 -- an example of an event with well sampled nights, which are clearly seen as the seven vertical lines of data at $\sim$constant $t$ ($t\propto x/R_{\rm E}$) in the residuals plot. They can be seen less clearly in the light curve plot.}
\protect\label{mn269}
\end{figure}

To judge the validity of the stated (and adjusted) error bars, the scatter in photometric data was investigated. There exists within the data some 14 nights of observation when more than 10 individual photometric measurements of the same source were acquired in a night's observation (e.g. 2002-BLG-269 -- Fig. \ref{mn269}). For events where there is a night of intensive observation away from the peak, the light curve would not be expected to change much over the course of the night. Therefore the scatter in the data points taken on these nights can be used to judge errors. 

\begin{figure}
\begin{center}
\includegraphics[angle=0,width=0.47\textwidth]{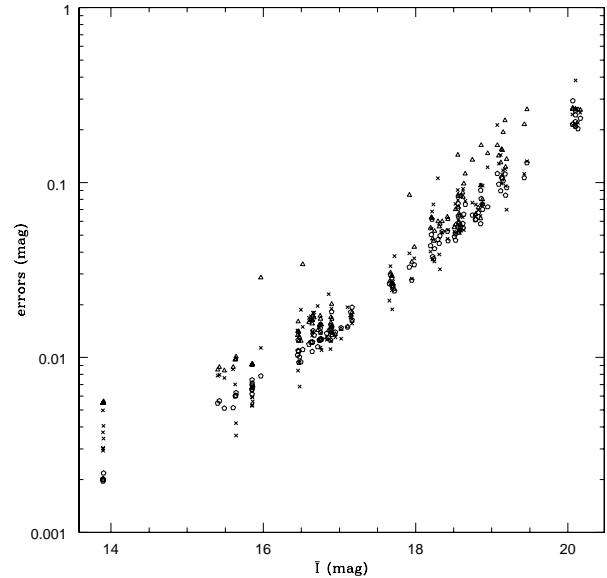}
\end{center}
\caption{RMS Scatter ($S_i$ -- crosses), reported errors ($\sigma_i$ -- circles), and adjusted error bars ($s_i$ -- triangles), compared to the mean magnitudes of the detections taken that night, $\bar{I}$. The scatter clearly increases for fainter sources, as expected, due to the lower signal-to-noise ratio at lower brightness. The scatter and reported error bars are largely consistent, with scatter generally greater than the reported error bars, but within the scaled error bars.}
\protect\label{mag-scatter}
\end{figure}

\begin{figure}
\begin{center}
\includegraphics[angle=0,width=0.47\textwidth]{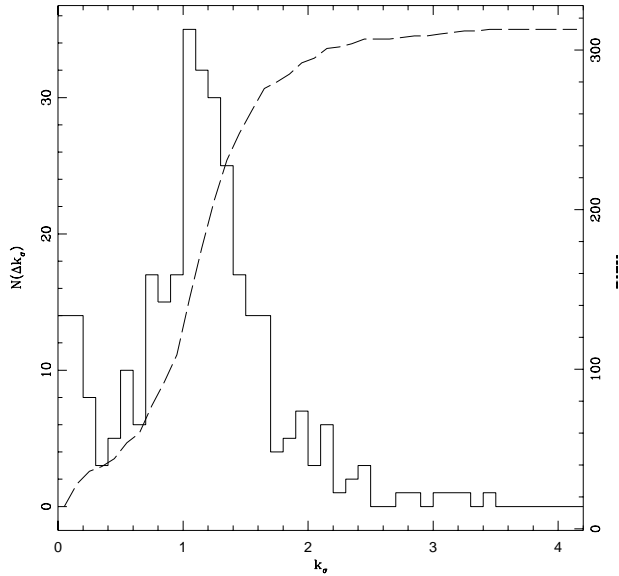}
\end{center}
\caption{Histogram of values of $k_\sigma$, the error bar scaling factor. The dashed line shows the cumulative histogram. It can be seen that the value for $k_\sigma$ is most commonly found in the range 1.1 -- 1.3, which implies that the reported error bars are smaller than the actual error, according to these 7-parameter fits. The median value of $k_\sigma$ found here is in excellent agreement with the scaling factor found in Section \ref{scatter}, which implied that actual errors where $\sim 1.2$ times the reported errors.}
\protect\label{fhist}
\end{figure}

Data were taken from the 128 nights, in 22 events, where 10 or more frames were observed in a single night. The photometric scatter, $S_i$, was determined for each by finding the root-mean-squared deviation from the mean $\bar{x}$ of the magnitude data $x$:
\begin{equation}
S_i^2=\frac{1}{N-1} \sum_{j=1}^N (x_j - \bar{x})^2,
\end{equation}
where there are $N$ data points on the night $i$. This RMS scatter is a reasonable way to judge the errors in the photometry, as \Code{PLENS} assumes a Gaussian distribution of errors. $S_i$, $\sigma_i$ and $s_i$ correlate very well; the scatter observed is consistent (Fig. \ref{mag-scatter}) with the mean reported error bar for the nights studied -- scaling directly by a factor of $\sim$1.2. A histogram of the values of $k_\sigma$, the error bar scaling factor, (Fig. \ref{fhist}) fitted to the 321 analysed events, agrees with this hypothesis -- it peaks around 1.2. 

The sample of 22 events that contain these well-sampled nights represents 7 percent of the 2002 events studied. They are spread throughout the data; it would be a reasonable assumption to extend the conclusions of this scatter analysis to all events. This would indicate that the scatter in data in other events is approximately the same as the quoted photometric error. $S_i$ and $s_i$ also correlate strongly -- with the adjusted error bars consistently larger. We concluded that the scaling of error bars by \Code{PLENS} is sufficient to encompass any likely photometric scatter, which should prevent any noise appearing as a planetary candidate.

\section{
2002 OGLE events
}\label{2002data}

\subsection{
Identification of suitable events
}\label{reject}
We selected events from the OGLE data that were reasonable approximations to a PSPL model. As the OGLE EWS\footnote{\texttt{http://www.astrouw.edu.pl/$\sim$ogle/ogle3/ews/ews.html}}  system looks for any brightening which could be a lensing event, it occasionally puts out a false alert -- the `event' later turns out to be a variable star, for example. Also, a number of the events observed have high mass ratio binary lenses (i.e. a binary star acting as a lens), which \Code{PLENS} is not designed to deal with. The non-PSPL events were removed from the sample. 

These events were identified by assessing the accuracy of the fitted parameters for each event. A Monte-Carlo method was used to produce data sets based on the OGLE data, with new values for each point chosen within a Gaussian distribution based on the stated error bars. The RMS variation in the parameters fitted to these data sets then provided an uncertainty measurement on each parameter. Fits without blending were carried out on 200 fake data sets for each event, providing error bars with a $1/\sqrt{200} \approx 7$ percent uncertainty. 

It was found that the baseline magnitude $I_0$ and peak time $t_0$ were well determined for most events, while the amplification $A_0$ and event timescale $t_{\rm E}$ can vary considerably in events without a clearly defined microlensing light curve. The uncertainties in these parameters, $\sigma_{A_0}$ and $\sigma_{t_{\rm E}}$, were used to provide filters to remove poor events. A sieve of four filters was used to catch non-PSPL events:

\begin{enumerate}
\item The three events listed as mistakes on the OGLE web page were removed.
\item Events with $\sigma_{A_0} > \frac{1}{2} A_0$ or $\sigma_{t_{\rm E}} > \frac{1}{2} t_{\rm E}$ were removed -- i.e. events with over 50 percent uncertainty in either of these parameters.
\item If an event's $\sigma_{A_0}$ and $\sigma_{t_{\rm E}}$ were greater than 3 times their respective median values, the event was removed. This caught events that did not contain enough data to constrain these parameters, for example those only observed during the rising part of the light curve. 
\item A brief visual inspection caught those events with clear non-PSPL light curves, that had passed through the other filters.
\end{enumerate}

From the original list of 389 events the filters rejected 3, 42, 10 and 13 events respectively  -- a total of 68, which are listed in Appendix \ref{rejected}. This left 321 events that were then searched for planetary candidates.

\subsection{
Planetary candidates
}\label{candidates}

A web page interface simplified the search for planetary candidates. By listing the events in order of decreasing $\Delta\chi^2$ -- with $\Delta\chi^2$ now representing the `best' value from each map; the magnitude of $\Delta\chi^2$ where it is at its most negative, and the planet model is most likely to be correct -- the events with the largest discrepancy from PSPL could be easily identified.  Examination of the original photometric data, the $\Delta\chi^2$ maps, and \Code{PLENS}'s output of the \mbox{$x$-$y$} grid position of the best point, allowed the data point corresponding to the `planet' to be identified. 

Of the analysed sample of 321 events, the 12 events listed in Table \ref{table-candidates} have $\Delta\chi^2 > 25$. These high $\Delta\chi^2$ events do not necessarily represent planetary candidates. Examination of the individual frames which gave the planet-like anomalies (Udalski, private communication) revealed that 10 of these events have anomalies due to other factors, such as clouds etc., as detailed in Table \ref{table-candidates}.

\begin{table}
\caption{Data points with $\Delta\chi^2 > 25$ following a fit including blending with \mbox{$q=10^{-3}$}. The reason for each anomaly is listed in the last column; those marked with an asterisk were suggested by Udalski (private communication). Those events listed as planetary candidates are the events for which the observation in question was OK, and no alternative explanation was given for the anomaly.}
\protect\label{table-candidates}
\centering
\begin{tabular}{|l|l|r|p{0.15\textwidth}|}
\hline
\rule{0pt}{2.5ex}Event & Frame & $\Delta\chi^2$ & Anomaly\\
& (HJD-2450000) & & \\
\hline
2002-BLG-140 & 2434.81238 & 110.03  & Planet candidate?\\
2002-BLG-055 & 2424.89865 &  64.97  & Planet candidate.\\
2002-BLG-205 & 2472.50459 &  48.06  & Clouds.* \\
2002-BLG-110 & 2472.50459 &  47.53  & Clouds.* \\
2002-BLG-085 & 2426.74382 &  43.63  & Clouds.* \\
2002-BLG-321 & 2517.62203 &  34.60  & Lens is a stellar mass binary.* \\
2002-BLG-009 & 2143.52890 &  32.29  & Lone high data point well before peak. Data taken in 2001 observing season, early in OGLE-III when photometry was not as accurate. \\
2002-BLG-186 & 2472.65734 &  32.01  & Clouds.* \\
2002-BLG-011 & 2410.74028 &  28.54  & Clouds.* \\
2002-BLG-311 & 2472.65734 &  28.50  & Clouds.* \\
2002-BLG-156 & 2471.54596 &  28.15  & Bad image - highly out of focus.* \\
2002-BLG-065 & 2453.54649 &  25.44  & Clouds.* \\
\hline
\end{tabular}
\end{table}

Of the two remaining events, we regard 2002-BLG-055 as a good planetary microlensing candidate, and 2002-BLG-140 as a possible candidate.

\begin{figure}
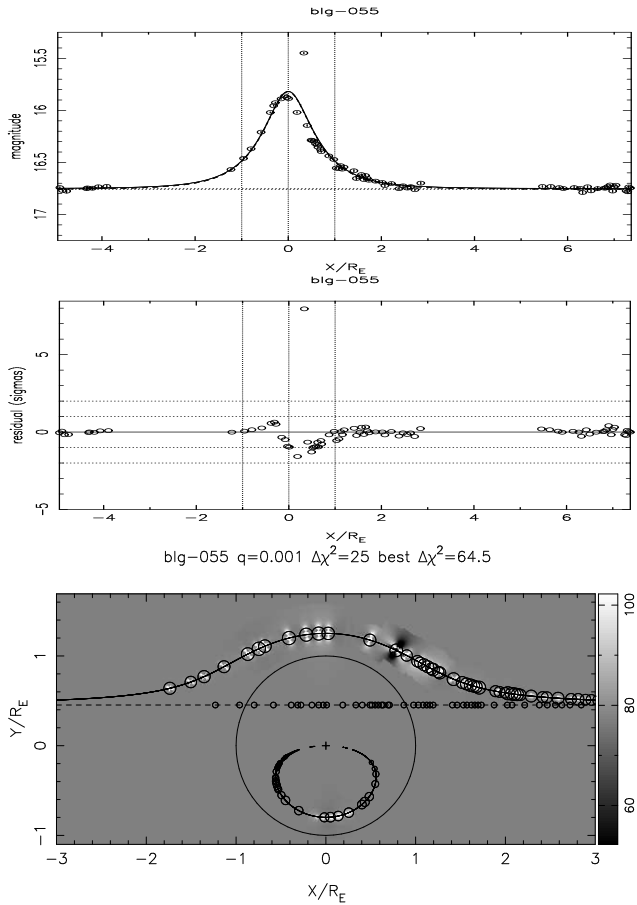

\begin{center}
\includegraphics[angle=-90,width=0.47\textwidth,totalheight=0.2\textwidth]{fig7a.ps}\\
\includegraphics[angle=-90,width=0.47\textwidth,totalheight=0.2\textwidth]{fig7b.ps}\\
\includegraphics[angle=270,width=0.47\textwidth]{fig7c.ps}
\end{center}
\caption{Light curve and $\Delta\chi^2$ map for 2002-BLG-055, which is the most likely planetary candidate as the strong detection shown here is in agreement with independent work.}
\protect\label{chi055}
\end{figure}

\begin{figure}
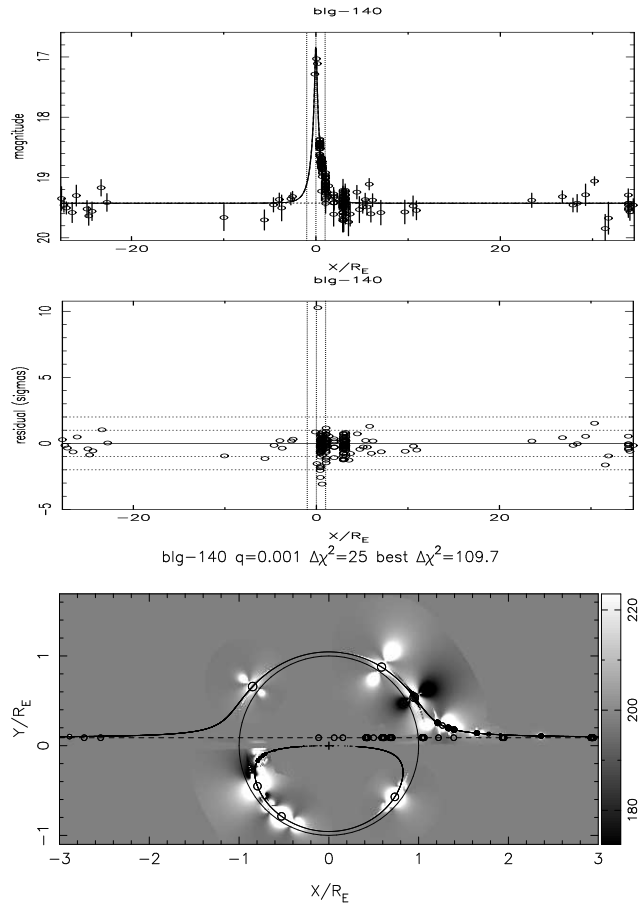

\begin{center}
\includegraphics[angle=-90,width=0.47\textwidth,totalheight=0.2\textwidth]{fig8a.ps}\\
\includegraphics[angle=-90,width=0.47\textwidth,totalheight=0.2\textwidth]{fig8b.ps}\\
\includegraphics[angle=270,width=0.47\textwidth]{fig8c.ps}
\end{center}
\caption{Light curve and $\Delta\chi^2$ map for the second possible planetary candidate, 2002-BLG-140. The data deviates strongly from a PSPL model; the peak $\Delta\chi^2=110$. This is clearly visible as a dark detection zone after the peak.}
\protect\label{chi140}
\end{figure}

\subsubsection{2002-BLG-055}
One very high data point shortly after the peak of this well sampled event appears to be due to a binary lens with $q$ in the expected range for a star and massive planet system (Fig. \ref{chi055}). This point was also judged to be a good planetary candidate by \citet{JP}. They find a good fit to the data using a parallax model, with three more fitted parameters, although the result from this more sophisticated model does not differ greatly from the PSPL model presented here (see Table \ref{blg055table}). 
The anomalous data point is around 0.6 mag higher than the fitted PSPL curve, with errors on this event at $\sim0.01$ mag (recorded) and $\sim0.1$ mag (adjusted). This is a very strong candidate for the detection of a planet; the fit suggests a planet with a projected orbital radius of around $a=1.3R_{\rm E}$. A brief survey over $q$ shows that this single deviating point is not sufficient to uniquely determine a fit - a range of mass ratios between 0.001 and 0.01 are equally likely.

\begin{table}
\caption{Comparison of fitted parameters to event 2002-BLG-055 found by our PSPL model, to those found by Jaroszy\'nski and Paczy\'nski (parallax model). Data taken from \citet{JP}.}
\begin{center}
\begin{tabular}{|l|c|r|}
\hline
PSPL & & parallax \\
\hline
2452406.3 $\pm$ 0.2 & $t_0$ & 2452401 \\
110.8 $\pm$ 0.4 & $t_{\rm E}$ & 110 \\
16.751 $\pm$ 0.001 & $I_0$ & 16.75 \\
2.370 $\pm$ 0.002 & $A_0$ & 2.249 \\
\hline
\end{tabular}\\
\end{center}
\protect\label{blg055table}
\end{table}

\subsubsection{2002-BLG-140}
This event has the largest $\Delta\chi^2$ of all those selected; it is the only event to have $\Delta\chi^2$ greater than the $\Delta\chi^2_T=100$ threshold, roughly corresponding to a 10$\sigma$ deviation. A dark zone on the $\Delta\chi^2$ map (Fig. \ref{chi140}), at the image position of the second data point after the peak, shows that the $q=10^{-3}$ binary model clearly provides a better fit to the data than a PSPL model for this event. A check on the CCD frame responsible for the image confirmed that the observation was OK, so the point is valid -- there are no defects to explain the anomaly as non-planetary. However this is a high amplification event ($A_0=11.3$), and the anomalous point is one of only three around the peak. Without a higher sampling rate at this critical point, and on the rising part of the light curve, it is possible to fit a number of models to the data, so we can only regard this event as possible planetary microlensing. A preliminary search of a parameter-space involving $q, x$, and $y$ revealed that the highest $\Delta\chi^2$ is found with a mass ratio of 0.01.  The $x,y$ co-ordinates of the 'planet' remained approximately constant throughout, around the image position of the $q=10^{-3}$ anomaly at (1.0,0.58) -- corresponding to a projected orbital radius of $a=1.16$. This is near to $R_{\rm E}$, as expected for a candidate with such a strong influence on the event.

\subsection{Lens masses}
Following the method of \citet{THK} a lens mass can be estimated from the event timescale $t_{\rm E}$; the best fit mass ratio can give a likely mass for the planetary candidates. The two events, 2002-BLG-055 and 2002-BLG-140 have $t_{\rm E} = 110.81$ and 22.28, corresponding to lens masses of 0.53 and 0.24 ${\rm M}_\odot$ respectively. The mass ratios then give masses of the 'planets' as approximately between 0.6 and 6 ${\rm M}_{{\rm Jup}}$ for 2002-BLG-055, and 2 ${\rm M}_{{\rm Jup}}$ for 2002-BLG-140. The projected orbital radii, $a$, for these candidates are 8.6 and 1.5 AU -- if there is a planet around the 2002-BLG-055 lens, it is at a distance greater than radial velocity searches can currently probe.

\section{
The abundance of cool planets
}\protect\label{ptot}

The single good planet-like anomaly in the OGLE-III data places a significant constraint on the abundance of planets. However, the OGLE-III data are sparsely sampled in time, so that many planet-like anomalies can be missed in the gaps between data points.  We can, however, calculate the expected number planet-like anomalies that would be detected in this dataset, for various assumptions about the planet abundance.

\begin{figure}
\begin{center}
\includegraphics[width=0.47\textwidth]{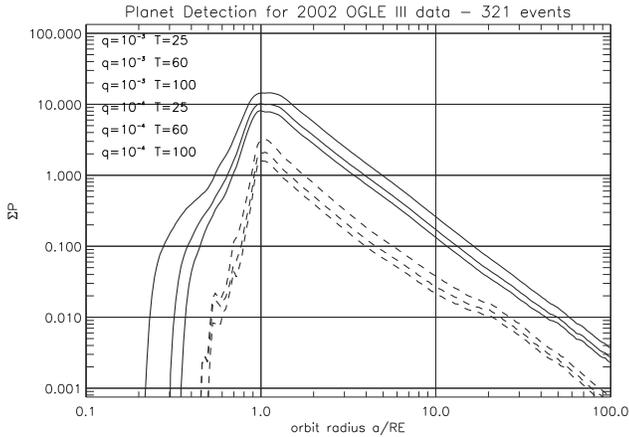}
\end{center}
\caption{Sum of detection probability, at $\Delta\chi^2 > 25, 60$ and 100 thresholds, of planets with $q=10^{-3}$ and $10^{-4}$, as a function of orbital radius in $R_{\rm E}$. Solid lines --  $q=10^{-3}$, dashed lines -- $q=10^{-4}$.}
\protect\label{ptotplot}
\end{figure}

The total number of detections expected is found by summing the individual detection probabilities at each orbital radius -- with $a$ in units of $R_{\rm E}$, so the detection zones do not need to be scaled for lens stars of different masses (Fig. \ref{ptotplot}). This was done for mass ratios of $10^{-3}$ and $10^{-4}$ and $\Delta\chi^2$ thresholds of $\Delta\chi^2_T = 25, 60$, and 100. For isolated data points, the detection probability scales with planet mass $m$, so the sum of detection probability for the lower mass ratio companions is considerably lower. The peak values of the sum of probabilities, and total number of candidates found, are displayed in Table \ref{limitsum}.

We concentrate primarily on the candidates and detection probabilities at $\Delta\chi^2_T = 25$, $q=10^{-3}$, with a fit including blending. The summed probability peak of $\sim 14$ (Fig. \ref{ptotplot}) implies that the expected tally of planetary candidates should be 14, if every star has a planet with that mass ratio. The peak probability was found to be at $a \approx R_{\rm E}$, as expected.

\begin{figure}
\begin{center}
\includegraphics[width=0.47\textwidth]{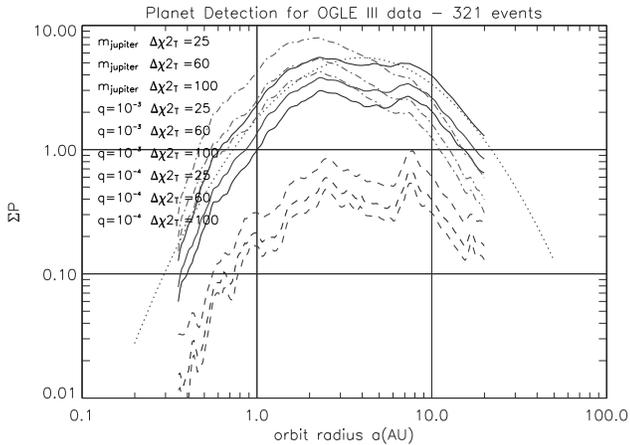}
\end{center}
\caption{Summed probability plot, with $a$ expressed in AU. Solid lines --  $q=10^{-3}$, dashed lines -- $q=10^{-4}$, dot-dashed lines -- extrapolation to $m={\rm M}_{\rm Jup}$. For each set, the number of expected detections is plotted against orbital radius for $\Delta\chi^2_T > 25, 60$ and 100. The parabolic shape of the curve is clear; an empirical fit (Equation \ref{sigPeqn}) to it gives an expression for the population of planets in terms of $a$, $q$, $n$ and $\Delta\chi^2$ -- dotted line.}
\protect\label{probreal}
\end{figure}

The expected number of planets per star is given by:
\begin{equation}\label{limits}
<n_p>=\frac{n}{\Sigma P},
\end{equation}
where $n$ is the number of candidates, and $\Sigma P$ represents the peak value of the summed probability plots (although this result is intuitive, its derivation is non-trivial, and is given in Appendix \ref{proof}). The $q=10^{-3}$ analysis gives $n=1$, or possibly 2 if 2002-BLG-140 is regarded as a candidate, and $\Sigma P=14.5$.  This gives $<n_p>=0.07$, implying that not more than 1 in 14 lens stars have planets with $q = 10^{-3}$, and $a \approx 1 R_{\rm E}$.

To look for a higher confidence level of planet detection, the analysis was repeated for $\Delta\chi^2_T=60$ and 100. Only 2002-BLG-140 has a $\Delta\chi^2$ above the highest level. The predicted number of detections is also lower; $\Sigma P \approx 10$ and 8 for $\Delta\chi^2_T=60$ and 100, respectively. 

\begin{table}
\centering
\caption{Summary of upper limits on galactic planet population, where $<n_p>$ is the expected number of planets per star, at  $a=R_{\rm E}$, given $n$ detections where $\Sigma P$ were expected, for various values of mass ratio $q$ and $\Delta\chi^2$ threshold, $\Delta\chi^2_T$. The figures in brackets are the values found if 2002-BLG-140 is taken as a planetary candidate.}
\protect\label{limitsum}
\vspace{5mm}
\begin{tabular}{|c|c|c|c|c|}
\hline
\rule{0pt}{3ex}$q$ & $\Delta\chi^2_T$ & $\Sigma P$ & $n$ & $<n_p>$\\
\hline
$10^3$ & 25 & 14.50 & 1 (2) & 0.07 (0.14) \\
$10^3$ & 60 & 10.19 & 1 (2) & 0.10 (0.20) \\
$10^3$ & 100 & 8.15 & (1) & (0.12) \\
\hline
$10^4$ & 25 & 3.21 & 1 (2) & 0.31 (0.62) \\
$10^4$ & 60 & 2.11 & 1 (2) & 0.47 (0.95) \\
$10^4$ & 100 & 1.63 & (1) & (0.61) \\
\hline
\end{tabular}
\end{table}

Our analysis for $q = 10^{-4}$ yielded the same candidates with $\Delta\chi^2 > 25$, and a sum of all detection probabilities with a peak value of 3.2.  This gives an upper limit on the population of less massive planets at 0.31 -- implying that smaller planets are much more common; 30\% of the lens star in this sample could have lower mass companions. This confirms the expected result from the d$N/$d$m \propto m^{-1}$ planet mass function observed in results from radial velocity searches \citep{MB}.

Following \citet{THK}, we translate our results from dimensionless variables ($a$/$R_{\rm E}, q$) into physical variables ($a$/AU, $m$/M$_{\rm Jup}$) using:
\begin{equation}
R_{\rm E} = 1.9\textrm{AU}\left(\frac{t_{\rm E}}{35 \textrm{d}}\right),
\end{equation}
and
\begin{equation}\label{fixedbug}
M = 0.3{\rm M}_\odot \left(\frac{t_{\rm E}}{35 \textrm{d}}\right)^2.
\end{equation}
It should be noted that Equation \ref{fixedbug} replaces Equation 17 of \citet{THK}, which was found to contain an error.

Fig. \ref{probreal} shows that $\Sigma P({\rm det}|a,q,\Delta\chi^2)$ is well approximated by:
\begin{equation}
\Sigma P \approx 5.5 \frac{q}{10^{-3}}\left(\frac{\Delta\chi^2}{25}\right)^{-1/2} \exp{\left\{-\frac{1}{2}\left[\frac{\log{(a/4{\rm AU})}}{0.4}\right]^{2}\right\}},
\protect\label{sigPeqn}
\end{equation}
which captures the scaling of $\Sigma P$ with $q$ and $(\Delta\chi^2)^{-\frac{1}{2}}$, and the approximately log-normal distribution of event timescales. Using Equation \ref{limits}, this gives an empirical result predicting the number of planets per star as a function of $a$, $q$ and $\Delta\chi^2$, given an observational constraint on $n$ -- the number of planets detected:
\begin{equation}
<n_p>\approx 3.6\times 10^{-5}\frac{n\sqrt{\Delta\chi^2}}{q} \exp{\left\{3.2\left[\log{\left(\frac{a}{4{\rm AU}}\right)}\right]^{2}\right\}}.
\end{equation}

Conversion to physical units (M$_{\rm Jup}$, and AU) broadens and lowers the peak in detection capability (Fig. \ref{probreal}), as the range in lens star masses gives a range in $R_{\rm E}$ within the sampled events. We find a peak in sensitivity at $a\approx4$~AU of 5.5 expected detections; this gives a limit on `cool Jupiters' ($q=10^{-3}$,$a\approx4$~AU) of $n_p \approx n/5.5 = 18$\%, assuming one detection.

\section{
Comparison with previous work
}
The large sample (321 events) studied allows significant upper limits to be placed on the galactic planetary population. The result presented here, a limit of $7n$ percent, is considerably stronger than the limits previous years' microlensing data could provide. \citet{THK} analysed 145 events, in three years (1998--2000) of OGLE-II data, and placed the upper limit at $21n$ percent, for detection of planets with $q=10^{-3}$ at a $\Delta\chi^2_T=25$ level. As the method for assessing detection probabilities in the OGLE-II data was the same as the one employed by this paper, combination of these results was possible; a value of $\Sigma P \approx 22$ was found for all four years of OGLE-II \& III (1998--2000, 2002). This gives an upper population limit of $5n$ percent, for the values of $q$ and $\Delta\chi^2_T$ given above.

Results presented by the PLANET collaboration \citep{Gaudi}, based on the study of 42 intensely monitored events, gave a 33 percent limit on Jupiter mass planets between 1.5 and 4 AU, at a $\Delta\chi^2_T=60$ level. Looking at our result for `cool Jupiters' at a $\Delta\chi^2_T=60$ level, we find a peak at around 4 AU of $\approx3.5$, corresponding to an abundance of planets of $28n$ percent. The results presented are in very good agreement with previous limit estimates, improving on them due to the larger sample of events studied. These limits can only be strengthened by inclusion of the more than 400 events detected in the 2003 OGLE season, which are currently under analysis.

\section{
Summary
}\label{conclusion}
We analysed the 2002~OGLE-III observations of Galactic Bulge microlensing events to assess their capability to detect lightcurve anomalies arising from lensing by `cool' planets. From 389 events we select 321 that satisfy criteria for adequate coverage and compatibility with a point-source point-lens light curve model. Our model includes 2 parameters that adjust the published error bars to improve their representation of the fit residuals in both faint and bright parts of the light curve. We then identify 12 events that appear to have significant planet-like anomalies, with a detection threshold $\Delta\chi^2>25$. Most of these are attributable to identifiable causes (e.g. photometry degraded by clouds), leaving $n=1$ event, 2002-BLG-055 with $\Delta\chi^2=65$, which we regard as a good planetary lensing candidate. 2002-BLG-140, with $\Delta\chi^2=110$, is also a possible candidate. Our analysis of detection efficiencies indicates that the 321 selected OGLE-III events should reveal 14 planets, with a detection criterion $\Delta\chi^2 > 25$, if each of the lens stars has 1 planet with a Jupiter-like mass ratio $q=10^{-3}$ and orbit radius $a=R_{\rm E}$. This implies an abundance of these planets at $n_p \approx n/14$ = $7n$\%. Approximate allowance for the range in $R_{\rm E}$ from event to event spreads the detection probability over a wider range of $a$, and reduces the peak sensitivity to 5.5 detections for planets with $a\approx4$~AU. This gives the abundance of `cool Jupiters' ($q=10^{-3}$,$a\approx4$~AU) per lens star as $n_p \approx n/5.5$ = $18n$\%. The detection probability scales roughly with $q$ and $(\Delta\chi^2)^{-1/2}$, and drops off from the peak at $a\approx4$~AU like a Gaussian with a dispersion of 0.4 in $\log{a/4~AU}$ (see Fig.\ref{probreal}). The peak in sensitivity at $a\approx4$~AU shows that microlensing is a complimentary method to other planet search techniques, which are more sensitive to close `hot Jupiters'. The anomalies we identify are confined to single data points in the OGLE-III data. Higher time-resolution monitoring of more events is required to verify the above measurement of the abundance of `cool Jupiters', and to characterise the duration of the anomalies in order to assess the masses of the planets involved.

\section{
Acknowledgements
}
We would like to thank the OGLE team for making their data public, and especially Dr.~A.~Udalski for checking the validity of the candidate events identified. We would also like to thank Martin Dominik for his helpful comments and suggestions.

\appendix
\section{Proof of upper limit equation}
\label{proof}

The equation for upper limits on planet population (\ref{limits}) is found by using Bayes theorem and Poisson statistics to describe the probability of detecting planets based on a prior assumption about their distribution. The proof is given below:\vspace{3mm}\\
\vspace{1.5mm}
Definitions:\\
\vspace{1.5mm}
\begin{tabular}{l c p{0.35\textwidth}}
$n_p$&=&number of planets per star\\
$E$&=&$n_p \Sigma(P)$ \\
&=&expected number of detections summed over all events\\
$n$&=&actual number of planets detected over all events
\end{tabular}\\
Bayes theorem:
\begin{equation}
P(n_p|n) = \frac{P(n|n_p) P(n_p) }{ \int_0^{\infty} P(n|n_p) P(n_p){\rm d}n_p}
\end{equation}
Poisson counting statistics:
\begin{equation}
P(n|n_p) = \frac{E^n e^{-E}}{n!}
\end{equation}
Prior assumption on $n_p$:
\begin{equation}
P(n_p) \propto n_p^{-m}, 
\end{equation}
$m = 0$ for prior uniform in $n_p$ for $n_p>0$\\
$m = 1$ for prior uniform in $\log{(n_p)}$ for $n_p>0$\vspace{3mm}\\
Expected value of $n_p$ given $n$ detections:
\begin{equation}
<n_p|n> = \frac{\int_0^{\infty} n_p P(n_p|n) {\rm d}n_p}{\int_0^{\infty} P(n_p|n) {\rm d}n_p}
\end{equation}
\begin{equation}
<n_p|n> = \frac{\int_0^{\infty} n_p E^{n} e^{-E} P(n_p) {\rm d}n_p}{\int_0^{\infty} E^{n} e^{-E} P(n_p) {\rm d}n_p}
\end{equation}
Now $n_p = E/\Sigma(P)$ and d$n_p = {\rm d}E/\Sigma(P)$, so
\begin{equation}
<n_p|n> = \frac{1}{\Sigma(P)} \frac{\int_0^{\infty} E^{n-m+1} e^{-E} {\rm d}E}{\int_0^{\infty} E^{ n-m } e^{-E} {\rm d}E}
\end{equation}
integrating by parts:
\begin{equation}
\int_0^{\infty} x^{k} e^{-x} {\rm d}x = - x^{k} e^{-x} + k \int_0^{\infty} x^{k-1} e^{-x} {\rm d}x
\end{equation}
Conveniently, the first term vanishes when evaluated at the integration limits $x=0$ and $x=\infty$.
We therefore have:
\begin{equation}
\int_0^{\infty} E^{n-m+1} e^{-E} {\rm d}E = (n-m+1) \int_0^{\infty} E^{n-m} e^{-E} {\rm d}E
\end{equation}
and so:
\begin{equation}
<n_p|n> = \frac{(n-m+1)}{\Sigma(P)}
\end{equation}
A choice of $m=0$ implies an assumed prior on population of, for example, 1 planet per star. However this assumption gives the same assumed probability of $X < n_p < X+1$ for any number of planets per star, $X$ -- while lower values of $X$ are intuitively more likely. Taking $m=1$ implies an assumed probability that is uniform in $\log{(n_p)}$, which has the advantage that it automatically gives $n_p>0$. This gives the intuitive result:
\begin{equation}\label{f}
<n_p>=\frac{n}{\Sigma P},
\end{equation}
which states that the number of planets per star is given by the number found over the expected number.

\section{
Rejected events
}\label{rejected}

Events rejected from the analysis as being non-PSPL. The rejected events fall into one of four categories, as outlined below, depending on which criterion for being an acceptable PSPL fit each failed to meet. Remarks on each event from the OGLE web page are quoted directly, and marked with an asterisk.:\vspace{3mm}\\
\vspace{1.5mm}
\begin{center}
\begin{tabular}{|c p{0.35\textwidth}|}
\hline
\multicolumn{2}{|c|}{Rejection categories}\\
\hline
(i) & Event listed as a mistake on the OGLE web page.\\
(ii) & Event has over 50 percent error in $A_0$ or $t_{\rm E}$.\\
(iii) & Event's $\sigma_{A_0}$ and $\sigma_{t_{\rm E}}$ are greater than 3 times the median.\\
(iv) & Event has visibly non-PSPL light curve.\\
\hline
\end{tabular}\\
\vspace{1.5mm}\end{center}
\scriptsize
\begin{tabular}{|c|c|p{0.25\textwidth}|}
\hline
Event & Category & Comment \\
\hline 
2002-BLG-002 & (iii) & Only part of peak observed -- parameters cannot be uniquely determined.\\ 
2002-BLG-003 & (ii) & Variable star.* \\ 
2002-BLG-018 & (ii) & Variable star. Light curve appears parabolic in some sections. \\ 
2002-BLG-023 & (ii) & Variable star?* \\ 
2002-BLG-032 & (iii) & Only part of peak observed. \\ 
2002-BLG-040 & (iv) & Variable star?*  \\ 
2002-BLG-051 & (ii) & Double?*  \\ 
2002-BLG-057 & (ii) & Only part of peak observed. \\ 
2002-BLG-058 & (iii) & Only part of peak observed. \\ 
2002-BLG-059 & (iii) & Only part of peak observed. \\ 
2002-BLG-068 & (iii) & Double?* \\ 
2002-BLG-069 & (iv) & Binary according to PLANET team.* \\ 
2002-BLG-077 & (ii) & Cataclysmic Variable (I. Bond MOA group).* \\ 
2002-BLG-078 & (ii) & Peak defined by only one data point -- not enough data to constrain parameters. \\ 
2002-BLG-080 & (ii) & Variable star?* \\ 
2002-BLG-081 & (ii) & Variable star?* \\ 
2002-BLG-089 & (iv) & Variable star.* \\ 
2002-BLG-090 & (ii) & Variable star?* \\ 
2002-BLG-099 & (iv) & Double.* \\ 
2002-BLG-113 & (ii) & Variable star.* \\ 
2002-BLG-114 & (iv) & Double?* \\ 
2002-BLG-117 & (ii) & Second peak well separated from first. Wide binary? \\ 
2002-BLG-119 & (iv) & Double? Variable star?* \\ 
2002-BLG-120 & (ii) & High $A_0$ event with considerable variation in fitted parameters, even on subsequent runs with same data. \\ 
2002-BLG-127 & (ii) & Variable star.* \\ 
2002-BLG-128 & (ii) & Double? Definitely non-PSPL event. \\ 
2002-BLG-129 & (iv) & Double? Variable star?* \\ 
2002-BLG-135 & (iv) & Double?* \\ 
2002-BLG-143 & (ii) & Single? Double?* \\ 
2002-BLG-146 & (ii) & High $A_0$ event with considerable variation in fitted parameters. \\ 
2002-BLG-149 & (iv) & Severely blended.* Light curve suggests lens may be a binary star, as there is a visible double peak. \\ 
2002-BLG-151 & (iii) & Variable star?* \\ 
2002-BLG-152 & (iv) & Double? Variable star?* \\ 
2002-BLG-159 & (ii) & Variable star? Peaks throughout data. \\ 
2002-BLG-173 & (ii) & Very noisy, intrinsically faint, event without clear peak. \\ 
\hline
\end{tabular}

\begin{tabular}{|c|c|p{0.25\textwidth}|}
\hline
Event & Category & Comment \\
\hline 
2002-BLG-175 & (ii) & Variable star? Faint, with multiple peaks. \\ 
2002-BLG-176 & (iii) & Variable star? Faint, with multiple peaks. \\ 
2002-BLG-194 & (ii) & Only part of peak observed. \\ 
2002-BLG-196 & (ii) & Parameters vary considerably on different fits to this small amplification event. \\ 
2002-BLG-200 & (ii) & Variable star?* \\ 
2002-BLG-202 & (ii) & Variable star?* \\ 
2002-BLG-203 & (i) & Mistake!!!* \\ 
2002-BLG-206 & (ii) & Very noisy event without clear peak. \\ 
2002-BLG-209 & (ii) & Peak defined by only one data point. \\ 
2002-BLG-215 & (ii) & Variable star.* \\ 
2002-BLG-228 & (i) & Artefact from bleeding column* \\ 
2002-BLG-229 & (ii) & Variable star? No peak visible. \\ 
2002-BLG-255 & (i) & Mistake!* \\ 
2002-BLG-266 & (ii) & Any lensing event masked by excessive noise. No fit obtainable. \\ 
2002-BLG-272 & (ii) & Only part of peak observed. \\ 
2002-BLG-273 & (ii) & Variable star?* \\ 
2002-BLG-278 & (ii) & No clear peak. \\ 
2002-BLG-284 & (ii) & High $A_0$ event with considerable variation in fitted parameters. \\ 
2002-BLG-286 & (ii) & High $A_0$ event with considerable variation in fitted parameters. \\ 
2002-BLG-307 & (ii) & Not enough data to define peak. \\ 
2002-BLG-313 & (ii) & Only part of peak observed. \\ 
2002-BLG-316 & (iv) & Appears to be either a binary star lens or variable star. \\ 
2002-BLG-324 & (ii) & Only part of peak observed. \\ 
2002-BLG-334 & (iv) & Strong parallax effect -- very long $t_{\rm E}$, event still not complete (2003 season). \\ 
2002-BLG-360 & (iv) & Only part of peak observed. \\ 
2002-BLG-363 & (ii) & Only part of peak observed. \\ 
2002-BLG-380 & (ii) & Only part of peak observed. \\ 
2002-BLG-382 & (iii) & Only part of peak observed. \\ 
2002-BLG-383 & (iii) & Only part of peak observed . \\ 
2002-BLG-386 & (ii) & Only part of peak observed. \\ 
2002-BLG-387 & (iii) & Only part of peak observed. \\ 
2002-BLG-388 & (ii) & Only part of peak observed. \\ 
2002-BLG-389 & (ii) & Only part of peak observed. \\ 
\hline
\end{tabular}

\normalsize

\end{document}